\documentclass{ws-procs9x6}
\newcommand{\RA}{\mbox{$\rightarrow$}}

\newcommand{\mpl}{\mbox{$M_{\rm PL}$}}
\newcommand{\ada}{\mbox{$\dot{\alpha}/\alpha$}}
\newcommand{\daa}{\mbox{$\frac{\Delta \alpha}{\alpha}$}}

\newcommand{\bqa}{\begin{eqnarray}}
\newcommand{\eqa}{\end{eqnarray}}
\newcommand{\beq}{\begin{equation}}
\newcommand{\eeq}{\end{equation}}

\begin{document}

\title{TIME VARIATION OF FUNDAMENTAL CONSTANTS AS A PROBE OF NEW PHYSICS}

\author{PAUL LANGACKER}

\address{Department of Physics and Astronomy \\
University of Pennsylvania \\
Philadelphia, PA 10104 \\
E-mail: pgl@electroweak.hep.upen.edu \\
UPR-1036T}

\maketitle

\abstracts{Time variation of fundamental constants would not
be surprising in the framework of theories involving extra dimensions.
The variation of any one constant is likely to be correlated with
variations of others in a pattern that is diagnostic of the
underlying physics.}

\section{Introduction}
There has recently been reported evidence for a possible time
variation of the fine structure constant on cosmological time scales~\cite{webb}.
Such variations are not surprising in any theoretical framework for the
unification of basic forces involving extra dimensions or in which
dimensionless couplings are related to the expectation values of scalar fields.
However, the variation
of $\alpha$ is likely to be correlated with the variations in other fundamental
quantities, such as other gauge and Yukawa couplings, and the ratios of
such dimensionful scales as the unification and electroweak or supersymmetry-breaking
scales, or
the unification and gravity scales. Thus, the observation of such variations
is a powerful probe of the underlying physics. I briefly
summarize\footnote{Presented at the
{\it Dirac Centennial Symposium}, Florida State University,
Tallahassee, December 2002.}
relevant issues and describe an analysis and parametrization of these effects
done in collaboration with Matt Strassler and Gino Segr\` e~\cite{lss}.

\section{Theoretical motivations}
There have been speculations going back to the pioneering work of Dirac
in 1937 that the fundamental ``constants'' of nature
may vary in time~\cite{Dirac,examples,reviews,anthropic,anthropicrev}.
From a modern perspective,
time-variation is not surprising. For example, in superstring theories and many
brane-world scenarios, couplings
are associated with moduli (scalar fields), which could be time-varying.
In fact, time variation could be expected in any theory in which
some or all of the couplings are associated with the expectation values
of scalar fields~\cite{scalar,space}, 
provided that they vary on cosmological time
scales. In the standard model, for example, masses are proportional to the
expectation value of the Higgs field. Gauge and Yukawa couplings
can similarly be associated with the expectation values of scalar
fields that occur in higher-dimensional operators.

As a simple example, suppose there is a higher-dimensional
operator coupling a scalar $\phi$ to the electromagnetic tensor
$F_{\mu \nu}$,
\beq
 \mathcal{L}_{\rm elm} \sim  \frac{1}{4} \left[ 1 +
\frac{\lambda
\phi}{M_{\rm PL}} \right] F_{\mu \nu} F^{\mu \nu} + \cdots, \eeq
where $\lambda$ is dimensionless and \mpl \ is the Planck scale.
It is useful to then replace $A_\mu$ by $A^\prime_\mu$, where
\beq A_\mu = A^\prime_\mu \left( 1 - \frac{\lambda \phi}{ 2 
M_{\rm PL}} \right), \eeq
so that $A^\prime_\mu$ has a canonical kinetic energy.
The couplings of charged particles to $A^\prime_\mu$
will then be canonical in terms of a rescaled electric charge
$e^\prime$, related by 
\beq
e = e^\prime \left( 1  + \frac{\lambda \phi}{ 2 
M_{\rm PL}} \right).
\eeq
$e^\prime$ is universal, i.e., the rescaling is the same 
for all charged particles.
If $\phi$ were a constant classical field, then the effects of these 
rescalings would be unobservable. However, if $\phi$ varies with
time or in space, the effective electric charge $e^\prime$
would also vary. For example,
if $\phi$ is time dependent, it would satisfy
\beq
\ddot \phi + 3 H \dot \phi + \frac{\partial V}{\partial \phi} = 0, \eeq
where $H$ is the Hubble expansion rate and $V$ is the scalar potential.
$\phi$ could be associated with a field introduced for other 
purposes, e.g., quintessence~\cite{application},
or it might have no other cosmological significance (i.e., $\phi$ might
or might not contribute significantly to $H$).

In addition to the
time/space varation, there would be new operators associated
with the derivatives of $\phi$~\cite{newop}, which are usually assumed to be small 
for small variations\footnote{Bekenstein has recently argued that
they might in fact be relevant to the Webb et al. observations~\cite{newop2}.}.
There would also be new long-range forces
coupling to electromagnetic energy density  mediated
by the quantum of $\phi$. These would violate the equivalence
principle and could lead to strong
but model-dependent bounds~\cite{longrange,application}.

One objection to the notion of time varying couplings is that  in many frameworks
the natural scale for the rate of variation of, e.g., the fine structure constant
$\alpha$, might be expected to be 
\beq \dot \alpha / \alpha \sim M_{\rm PL} \sim 10^{+43}
s^{-1},
\eeq
while any actual variation is clearly very much smaller than this\footnote{This
is reminiscent of the flatness problem, expressed as the statement that
the natural time
scale for the evolution of the universe is $1/\mpl$ rather than 1.4 $\times 10^{10}$
yr.}. For example, the Webb et al. results suggest
\beq \dot \alpha / \alpha \sim 10^{-15}
yr^{-1}\sim 10^{-66} M_{\rm PL}.
\eeq
 It is tempting to assume that
since
$\dot
\alpha /
\alpha$ is so small compared to its natural scale it must be exactly zero or at least
unobservably small for some reason. However, it is worth considering  an analogy
with the cosmological constant: in
most frameworks the natural scale for the vacuum energy density, related
to the cosmological constant by $\rho_{\rm vac}= \Lambda_{\rm cosm} /8 \pi G_N$, is
$\rho_{\rm vac} \sim M_{\rm PL}^4$.  Most people assumed that since
$\rho_{\rm vac}$ is so much smaller than this, there must be some principle to
ensure $\rho_{\rm vac} = 0$. Recently, however, the Type IA supernova and CMB
data have independently indicated that
 \beq 
\rho_{\rm vac} \sim 10^{-124} M_{\rm PL}^4 \ne 0. \eeq
(The observed dark energy may not be a true cosmological constant. It could
be a time-varying quantity such as quintessence. For the purposes of this remark
it does not make any difference.)

If $\alpha$ does vary with time, then it is likely that other fundamental
constants, such as other gauge couplings $\alpha_i$, Yukawa
couplings $h$, the electroweak scale $v$, and the Newton constant $G_N =1/M_{\rm
PL}^2$ also vary\footnote{I
will take the view that only dimensionless couplings and ratios
of mass scales are physically meaningful, and that quantities
such as   $\hbar$   
and $c$ are derived quantities rather than
fundamental. In that case, they can be taken to be fixed at unity.
For a debate on such matters, see~\cite{debate}.} 
in a correlated way~\cite{lss,examples,cf,correlated}. The relation of
these quantities is presumably specified in any complete unified description
of nature, though the form of the relations depends on the theory. One
should therefore allow for the possibility that other quantities are varying
when interpreting the observational data. Observations
(or non-observations) of time or space variations can therefore be
viewed as a probe of the underlying physics and how the various
quantitites are related.

\section{Search for varying $\alpha$}
Webb et al.~\cite{webb} have studied the absorption of light from background
quasars  by molecular clouds in the redshift range $0.5 < z < 3.5$.
They apply a new ``many multiplet'' method to simultaneously study
many relativistic  (i.e., $O(\alpha^2,\alpha^4)$) splittings, 
obtaining evidence for an increase in $\alpha$,
\beq
\daa \equiv \frac{\alpha_z - \alpha}{\alpha}  =-(0.72\pm 0.18)\times 10^{-5},
\label{webbresult}
\eeq
where $\alpha_z \ (\alpha)$ refers to the fine structure constant
at redshift $z$ (at present).
This would correspond to $\ada \sim 10^{-15}/$yr for $\dot{\alpha}/\alpha =$
constant. Using a different method, Bahcall et al.~\cite{Bahcall} find
a result consistent with no variation, though with lower precision,
$\daa = (-2 \pm 1.2) \times 10^{-4}$, for the redshift range 0.16-0.80.
Similarly, Cowie
and Songaila~\cite{Cowie} constrain $X \equiv \alpha^2 g_p m_e/M_p$,
where the proton magnetic moment is $e g_p/2M_p$, from the 21 cm
hyperfine line in hydrogen at $z \sim 1.8$, and Potekhin {\it
et al.}~\cite{Potekhin} limit $Y \equiv M_p/m_e$ from molecular
hydrogen clouds at $z = 2.81$:
\beq
\frac{\Delta X}{X} = (0.7 \pm 1.1) \times 10^{-5}, \ \ \ \ 
\frac{\Delta Y}{Y} = (8.3^{+6.6}_{-5.0}) \times 10^{-5}.
\label{XYexp}
\eeq

There are also stringent laboratory limits\footnote{
Laboratory limits are reviewed in detail in~\cite{reviews}.} For example, Prestage et
al. obtain~\cite{prestage}
$ | \daa | < 1.4 \times 10^{-14}$ over 140 days, corresponding 
to $\ada < 3.7 \times 10^{-14}/$yr if constant. More
recently, Sortais et al. obtained~\cite{sortais}
$\ada < (4.2 \pm 6.9) \times 10^{-15}/$yr. Laboratory techniques may
ultimately be sensitive to $\ada < 10^{-18}/T$, where $T$ is the running
time~\cite{reviews}.

A very stringent limit comes from the OKLO natural reactor~\cite{reviews}.
In particular, the $^{149}Sm/^{147}Sm$ ratio is depleted by the
capture of thermal neutrons,
\beq
n + ^{149}Sm \RA ^{150}Sm + \gamma.
\eeq
The cross section is dominated by a very low energy resonance, involving
an almost exact cancellation between Coulomb and strong effects.
Thus, even a small change in $\alpha$ could be significant.
 This was analyzed by Damour and Dyson~\cite{dd}
who found that \ada \ is bounded to be between  $-6.7 \times 10^{-17}/$yr
and $+ 5.0 \times 10^{-17}/$yr, and by Fujii et al.~\cite{fujii},
who obtained  $\ada = (-0.2 \pm 0.8) \times 10^{-17}/$yr, both over
$2 \times 10^9$ yr. This is a very stringent result, but does not directly
contradict (\ref{webbresult}) because the latter  refers to an earlier
time period (around $(6-11) \times 10^9$ yr ago). Furthermore,
only the possible variation in $\alpha$ was considered in~\cite{dd,fujii}.
It is conceivable that the effects of varying $\alpha$ could have been cancelled
by a change in the strong interaction strength, $\alpha_s$.

Big Bang Nucleosynthesis, which occurred for redshift $\sim 10^9-10^{10}$,
implies~\cite{bbn} that $\daa < O(10^{-2})$, 
assuming that only $\alpha$ varies. This is weak compared with
(\ref{webbresult}) if \ada \ is constant in time, but could conceivably
be important if there were significantly enhanced effects at large redshift.

CMB results may eventually be able to constrain \daa \ at the $10^{-2}-10^{-3}$
level for $z \sim 1000$ from their effects on the ionization history of
the Universe~\cite{cmb}.

\section{Correlations with $\alpha_s, \ \ h, \ \ v, \ \ G_N, \cdots $}
If $\alpha$ varies with time, it is likely that other fundamental constants do also.
The correlations of their time dependences would be a probe of the
underlying theory of particle
physics~\cite{lss,examples,cf,correlated}.

For example, the observed low energy gauge couplings are consistent with the
unification of the running gauge couplings at a scale  $M_{G} \sim
3\times 10^{16}$ GeV,
predicted in simple supersymmetric grand unification~\cite{polonsky}:
\beq
\frac{1}{\alpha _{i}(M_{Z})}=\frac{1}{\alpha _{G}}+b_{i}t_{G}
\label{rgeeqn}
\eeq
where $\alpha_i, i=1,2,3$ are the gauge couplings associated with
 $U(1) \times SU(2) \times SU(3)$,
 $t_{G}=\frac{1}{2\pi } \ln \frac{M_{G}}{M_{Z}}\approx 5.32$, 
$\alpha_{G}^{-1}\approx 23.3$  is
the inverse of the common coupling at the unification scale, and the $b_{i}$ are the beta
function coefficients. In the MSSM, $b_{i}=\left( \frac{33}{5},1,-3\right).$
The (running) electromagnetic fine structure constant
is related by
$\alpha^{-1} = \frac{5}{3} \alpha_1^{-1} + \alpha_2^{-1} \sim 127.9$,
where all three couplings are evaluated at $M_Z$.
If gauge unification holds, either in the simple MSSM framework or
something similar, then it is likely that all three gauge couplings
will vary simultaneously~\cite{lss,cf}. 

The simplest possibility is that the dominant effect is a time
variation in $\alpha_{G}^{-1}$. In that case, it is straightforward
to show~\cite{lss} that the strong coupling $\alpha_s = \alpha_3$ has a
magnified variation,
\beq\label{alphasMZ}
\frac{\Delta \alpha _{S}}{\alpha _{S}}\simeq \frac{3
}{8}\frac{\alpha _{S}}{\alpha }\frac{\Delta \alpha }{\alpha }\sim 5.8\frac{
\Delta \alpha }{\alpha }
\eeq
where $\alpha_s$ is evaluated at $M_Z$ and we ignore the difference
in the relative variation of
$\alpha$ between scales 0 and $M_Z$. There is an even stronger variation in the QCD scale
$\Lambda_{\rm QCD}$, at which $\alpha_s$ becomes strong,
\beq\label{DeltaQCD}
\frac{\Delta \Lambda _{QCD}}{\Lambda _{QCD}}
\sim 34\frac{\Delta \alpha}{\alpha  },
\eeq
which is around $-25 \times 10^{-5}$ for the Webb et al. value (\ref{webbresult}).
This has a theoretical uncertainty (given the assumptions) of around 20\%.
Most hadronic mass scales (with the exception of the pion mass) are approximately
proportional to $\Lambda_{\rm QCD}$, so they are expected to have the same relative 
variation.

It is also reasonable to consider a variation in the electroweak scale $v \sim 246$
GeV (which sets the scale for $M_Z=g_W v$, where $\alpha_W =g_W^2/4\pi
=3 \alpha_1/5 + \alpha_2 $ ), or more
precisely in the ratio of 
$v$ to the unification
scale $M_G$~\cite{lss,examples,cf,correlated}. 
(Only dimensionless ratios of mass scales are
physically relevant, so we are implicitly measuring all masses with respect to $M_G$.)
In~\cite{lss} we define the phenomenological parameter $\kappa$ by
\beq
\frac{\Delta v}{v} \equiv \kappa \daa, \eeq
which implies that
\bqa
\frac{\Delta \alpha _{S}}{\alpha _{S}} &\sim & \frac{3
}{8}\frac{\alpha _{S}}
{\alpha }\left( 1 - \frac{10\alpha}{\pi} \kappa \right)\frac{\Delta \alpha}{\alpha }
\nonumber \\
\frac{\Delta \Lambda _{QCD}}{\Lambda _{QCD}}
&\sim &34 \left( 1 + 0.005 \kappa \right) \frac{\Delta \alpha}{\alpha  }.
\eqa
These corrections are small for $\kappa$ of order unity, but important for
larger $\kappa$. In fact, it is shown in~\cite{lss} that $\kappa \sim 70$
in theories in which $v$ is tied to the scale of soft supersymmetry breaking, and in which
supersymmetry breaking occurs in a hidden sector at a scale in which a (unified)
gauge coupling becomes strong! Even in this case, the correction to the $\Lambda_{\rm QCD}$
variation is only a factor of 1.35. 

It is useful to
introduce phenomenological parameters for the variation of other fundamental ``constants''.
In particular, the variation of the Yukawa coupling $h_a$ for fermion $a$
(so that its Higgs-generated mass is $m_a = h_a v$) is parametrized
as\footnote{The effects of the running of $h_a$ are described in~\cite{lss}.}
\beq \frac{\Delta h_a}{h_a} \equiv \lambda_a \daa. \eeq
Similarly, the variation of the Planck scale $\mpl = G_N^{-1/2}$
(again, only the ratio of \mpl \ to other masss scales is relevant)
is parametrized as
\beq \frac{\Delta \mpl}{\mpl} \equiv \rho \daa. \eeq
The possible variation of various observables can
 be expressed in terms of these parameters, and their values can in
principle be computed in any complete fundamental theory, allowing for
a more general treatment of time variation\footnote{It was argued in~\cite{cosm}
that a variation in $\alpha$ would upset the fine-tuned cancellations of
radiative corrections to the cosmological constant with other contributions,
with enormous effect. 
We take the view that such arguments are
not conclusive given our lack of understanding of why $\Lambda_{\rm cosm}$ is so small.}.
For example, for the quantities defined before
(\ref{XYexp}) one predicts\footnote{We ignore possible variations
in $g_p$ because it is well described in the constituent
quark model, where it is a Clebsch-Gordan coefficient.} the variations~\cite{lss},
\bqa
\frac{\Delta X}{X} & \sim & \left(-32 + \lambda + 0.8\kappa \right)
  \frac{\Delta \alpha}{\alpha} \sim (23 \pm 6) \times 10^{-5}  \nonumber \\
\frac{\Delta Y}{Y} & \sim & \left( 34 - \lambda - 0.8\kappa \right)
  \frac{\Delta \alpha}{\alpha} \sim (-24 \pm 6) \times 10^{-5},
  \label{XY}
  \eqa
where I have assumed a common value $\lambda$ for all the
Yukawa factors $\lambda_a$, and the numerical values are evaluated
using  $\lambda=\kappa=0$ and the Webb et al. value (\ref{webbresult}).
These are to be compared with the experimental results in (\ref{XYexp}).
Clearly, within this framework the observational results in (\ref{webbresult})
and (\ref{XYexp}) are consistent only if there is a delicate cancellation of effects, with
$\lambda + 0.8 \kappa \sim 32$. Other applications, including big bang nucleosynthesis,
the OKLO reactor constraints, and the triple $\alpha$ process, are
considered in~\cite{lss,examples,cf,correlated}.

\section{Conclusions}
\begin{itemize}
\item
Time (or space) variation of fundamental ``constants'' is plausible in any theory in
which they are dependent on the sizes or properties of extra dimensions, or
on other scalar fields.

\item The natural scale for such variations in many framewroks is
$\ada \sim \mpl \sim 10^{43}/s,$ which is very much larger than what
is allowed by observations.
However, it is at least possible that the true variations
are nonzero but very small for some reason, just as the vacuum energy is
much smaller than the natural scale of $\mpl^4$.

\item Webb et al.~\cite{webb} have reported a positive result
(\ref{webbresult}),
corresponding to
$  \dot \alpha / \alpha \sim 10^{-15}
yr^{-1}\sim 10^{-66} M_{\rm PL}$ for constant \ada.

\item If $\alpha$ varies, then it is possible that other
fundamental quantities, such as the other gauge couplings, Yukawa couplings,
or the dimensionless ratios of the electroweak, unification, and gravity
scales also vary in a correlated way that depends on the underlying physics.
Such variations should be allowed for in analyzing experimental/observational results,
and can in principle be a significant probe of the underlying physics.

\item The comparison between different classes of observations depends
on the time dependence of \ada, which in turn depends on
the type of scalar fields involved and their potentials.

\item There may be long-ranged forces associated with the time variation~\cite{longrange}.

\end{itemize}

\section*{Acknowledgments}
It is a pleasure to thank the conference organizers for support. 
Supported in part by Department of Energy grant DOE-EY-76-02-3071.

\end{document}